\documentclass[twocolumn,amsmath,amssymb,superscriptaddress,prx,floatfix,reprint,,longbibliography]{revtex4-2}
\usepackage{graphics,amssymb,amsmath,epsfig,color,textgreek}
\usepackage{amsthm}

\usepackage{graphicx}
\usepackage[dvipsnames]{xcolor}
\usepackage{todonotes}
\usepackage{dcolumn}
\usepackage{bm}
\usepackage[colorlinks=true,citecolor=cyan]{hyperref}
\hypersetup{colorlinks=true,citecolor=cyan,linkcolor=red,urlcolor=magenta}
\usepackage{braket}
\usepackage[normalem]{ulem}
\usepackage{cancel}
\usepackage{xfrac} 
\usepackage{amsmath}
\usepackage{amssymb}
\usepackage{amsthm}
\usepackage{svg}
\usepackage{lipsum}
\usepackage{hyperref}

\usepackage{algorithm2e}
\RestyleAlgo{ruled} 
\SetKwComment{Comment}{/* }{ */}

\newcommand{\ZZ}{\mathcal{Z}}

\usepackage[capitalise]{cleveref}
\crefname{section}{Sec.}{Secs.}
\crefname{table}{Tab.}{Tabs.}
\crefname{figure}{Fig.}{Figs.}
\crefname{definition}{Def.}{Defs.}
\crefname{lema}{Lem.}{Lems.}
\crefname{theorem}{Thm.}{Thms.}
\crefname{corollary}{Cor.}{Cors.}

\begin{document}

\title{Denoising of Imaginary Time Response Functions with Hankel projections}

\author{Yang Yu}
\email{umyangyu@umich.edu}
\affiliation{Department of Physics, University of Michigan, Ann Arbor, Michigan 48109, USA}

\author{Alexander F. Kemper}
\email{akemper@ncsu.edu}
\affiliation{Department of Physics, North Carolina State University, Raleigh, North Carolina 27695, USA}

\author{Chao Yang}
\email{cyang@lbl.gov}
\affiliation{Computational Research Division, Lawrence Berkeley
National Laboratory, Berkeley, CA 94720, USA}

\author{Emanuel Gull}
\email{egull@umich.edu}
\affiliation{Department of Physics, University of Michigan, Ann Arbor, Michigan 48109, USA}

\date{\today}

\begin{abstract}
Imaginary-time response functions of finite-temperature quantum systems are often obtained with methods that exhibit stochastic or systematic errors.
Reducing these errors comes at a large computational cost -- in quantum Monte Carlo simulations, the reduction of noise by a factor of two incurs a simulation cost of a factor of four.
In this paper, we relate certain imaginary-time response functions to an inner product on the space of linear operators on Fock space. We then show that data with noise typically does not respect the positive definiteness of its associated Gramian. The Gramian has the structure of a  Hankel matrix. As a method for denoising noisy data, we introduce an alternating projection algorithm that finds the closest positive definite Hankel matrix consistent with noisy data.
We test our methodology at the example of fermion Green's functions for continuous-time quantum Monte Carlo data and show remarkable improvements of the error, reducing noise by a factor of up to 20 in practical examples.
We argue that Hankel projections should be used whenever finite-temperature imaginary-time data of response functions with errors is analyzed, be it in the context of quantum Monte Carlo, quantum computing, or in approximate semianalytic methodologies.
\end{abstract}

\maketitle

\section{Introduction}

A fundamental task in the study of quantum field theories is the evaluation of
the expectation value of dynamical correlation functions in a statistical ensemble~\cite{altland2010condensed}.  For most non-trivial models, these correlation functions cannot be evaluated analytically, and one has to resort
to numerical methods of some form.  Chief among these are quantum
Monte Carlo (QMC) methods, which simulate field theories by stochastic sampling ~\cite{Blankenbecler1982,Hirsch1986,Prokofev98,Rubtsov2005,Werner06,Prokofev07,Bazavov10,Gull11RMP,Cohen15,Neuhauser16,Rossi17,Winograd20}.  These methods form the backbone of correlated lattice model \cite{Qin22} and materials simulations~\cite{gubernatis2016quantum} and are also used to study systems in high energy physics, such as quantum chromodynamics~\cite{rothe2012lattice,Bazavov10}. 
Many-body simulations of correlation functions on quantum computers are similarly of a statistical nature.

The predictive power of stochastic simulations of quantum field theories is often limited only by the precision to which correlation functions, such as Green's functions and susceptibilities, can be obtained. This precision is characterized by the size of the statistical error, which decreases only as the square root of the computational effort.   Thus, decreasing the uncertainty comes
at a high computational cost once some threshold in effort (and thus precision) has been reached.

The majority of statistical simulations of quantum field theories are performed in so-called imaginary time or Matsubara frequencies~\cite{Mahan}, though alternative numerically exact formulations based on Keldysh diagrammatics~\cite{Muhlbacher08,Werner09,Cohen15,Erpenbeck23} are possible on small systems. In the absence of a sign problem, this Wick-rotated formalism results in positive real simulation weights with a straightforward probabilistic interpretation. However, the interpretation of response functions requires an analytic continuation \cite{Jarrell1996,fei2021nevanlinna,Zhang23} that exponentially amplifies statistical and systematic uncertainties. 

Given the high computational cost, it is of great benefit to identify
components of the error that violate fundamental physical laws
or mathematical constraints. For example, if a quantum computing
simulation is known to have a fixed particle number, a sample that violates 
this constraint can be removed from consideration~\cite{charles2024simulating}.
Causality, or the positivity of the spectral function, is another such physical criterion. Noisy real-time simulations can make use of a causal projection based on an inner product structure and a projection to an associated positive semidefinite (PSD) Toeplitz matrix, leading to vastly improved spectra of time-evolved systems~\cite{Kemper23}.

This paper introduces a criterion based on an inner product structure  for equilibrium \emph{imaginary}-time correlation functions. 
We define an inner product of imaginary time ($\tau$)
operators on the Fock space and show its positivity. 
When discretized on a uniform grid, its Gramian matrix is a positive semidefinite Hankel matrix whose positivity may be violated in noisy simulations. We then devise an alternating projection algorithm to find the  positive-seminite Hankel matrix closest to given noisy data and thereby de-noise imaginary time data. In an application to
numerical results from typical QMC problems, we find that this projection substantially reduces the statistical error at negligible computational cost.

\section{Method}\label{sec:method}
For a system described with a Hamiltonian $H$ at inverse temperature $\beta$ and with density matrix $\rho = e^{-\beta H}/\ZZ$ and partition function $\ZZ=\text{Tr} (e^{-\beta H})$, the real-time correlation function $G(t,t')$ of an operator $A^\dagger(t)$ with another operator $B(t')$,

\begin{align}
    G_{AB}(t,t')=&\text{Tr}\left[\rho \left(e^{iHt}A^\dagger e^{-iHt}\right)\left(e^{iHt'}B e^{-iHt'}\right)\right],
\end{align}
induces an inner product $\langle A(t), B(t') \rangle$ on the vector space of linear operators $A(t), B(t')$ from the Fock space to itself, since it satisfies linearity, conjugate symmetry, and positivity when the operators
are identical: $G_{AA}(t,t) = \langle A(t), A(t) \rangle \ge 0$~\cite{Hyrkas22,Kemper23}.
The operators $A$ and $B$ may be either fermionic or bosonic.
As a consequence, for any sequence $(t_1, t_2, \cdots, t_N),$ and for any operator $A$, the Gramian $\underbar G_{ij}=G_{AA}(t_i,t_j)$, $1\leq i,j\leq N$ is a positive semidefinite $N \times N$ matrix that can be used as a starting point for denoising real-time simulations~\cite{Kemper23}. Moreover, if the correlation function is
time-translation invariant, i.e. $\underbar G(t_i,t_j) = \underbar G(t_i - t_j)$, the Gramian
matrix is a Toeplitz matrix.

\begin{figure}[tb]
    \centering
    \includegraphics[width=\columnwidth]{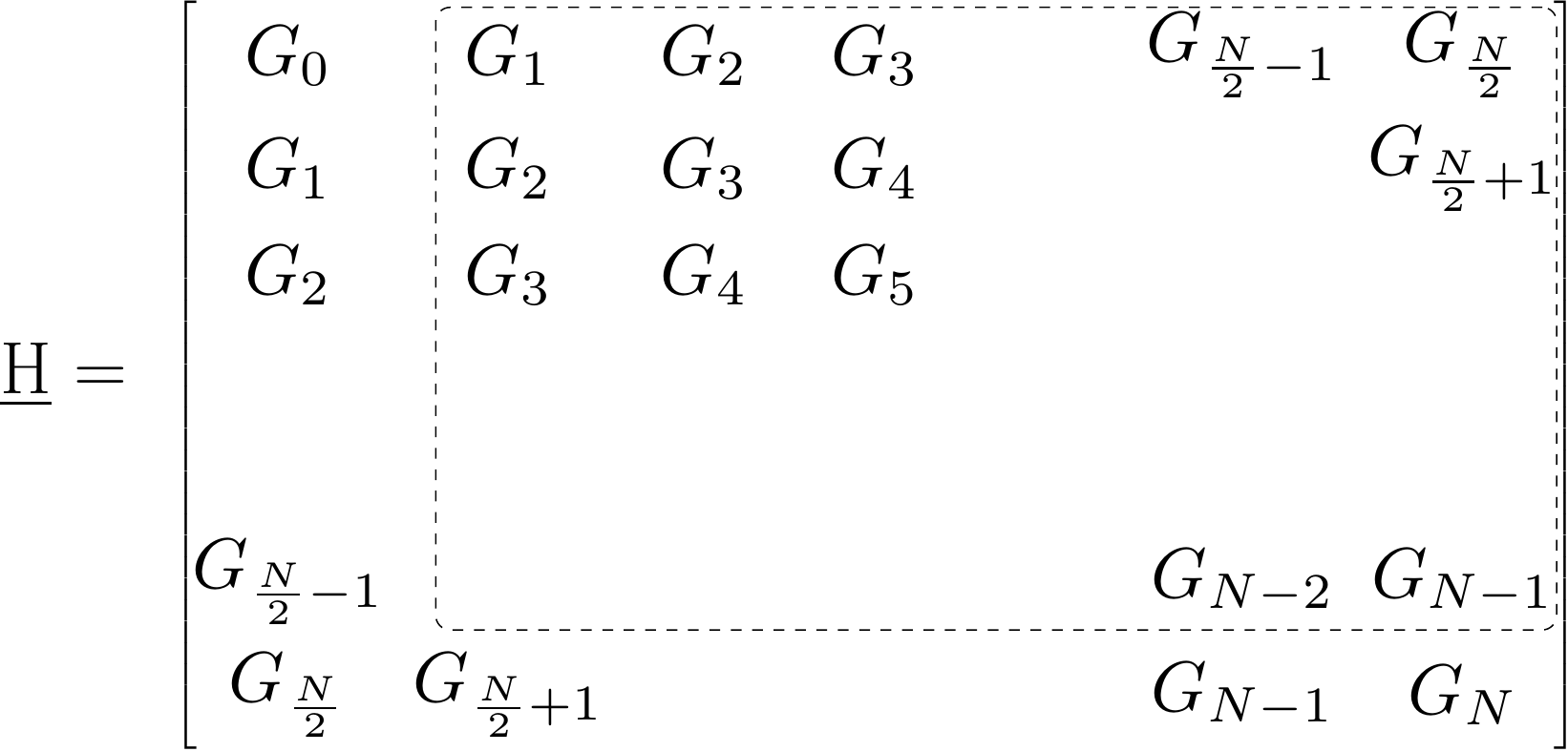}
    \caption{Hankel matrix $\underbar H_{ij}=G_{i+j}=G(\tau_i+\tau_j)$ for $i,j=(0, 1, \cdots, \frac{N}{2})$ used in the projection. Also shown (dashed line) is the Hankel matrix corresponding to $i,j=(\frac{1}{2}, \frac{3}{2}, \cdots,\frac{N-1}{2})$.
    }
    \label{fig:Hankel}
\end{figure}

Consider, in analogy, the imaginary time function
\begin{align}
    G_{AB}(\tau, \tau') = \text{Tr} \left[\rho \left(e^{\tau H}A^\dagger e^{-\tau H}\right)\left( e^{-\tau'H}B e^{\tau'H}\right)\right]
    \label{eq:lehmanntau}
\end{align}
This expression similarly satisfies linearity, conjugate symmetry, and for $0\leq \tau+\tau'\leq \beta$, $G_{AB}(\tau,\tau')$ coincides with the definition of an imaginary time correlation function of $\tau+\tau'$:
\begin{align}
  G_{AB}(\tau,\tau')=G_{AB}(\tau+\tau')=G_{AB}(\tau',\tau).
\end{align}
Due to the time-ordering in the definition of the imaginary-time correlation function, this coincidence is only true inside the interval $0 \le \tau+\tau' \le \beta$. 

The remaining requirement for $G_{AB}(\tau, \tau')$ to be an inner product, positivity, is straightforward.
Define $A(\tau)=e^{-\tau H}A e^{\tau H}$ with $A^\dagger(\tau)=e^{\tau H}A^\dagger e^{-\tau H}$, such that $G_{AA}(\tau,\tau')=\text{Tr}\left[\rho A^\dagger(\tau)A(\tau')\right]$. We note that $B(\tau)=A^\dagger(\tau)A(\tau)$ is positive semidefinite since it is a product of a matrix and its adjoint.
Thus, $G_{AA}(\tau,\tau)=\text{Tr}\left[ \rho B(\tau) \right] \geq 0$ for any $\tau$, since $\rho$ is also PSD, and $\text{Tr}[PQ]\geq 0$ for any two PSD matrices $P$ and $Q$. $G_{AB}(\tau,\tau')$ therefore forms an inner product 
$\langle A(\tau),B(\tau')\rangle$
on
the space of linear operators from the Fock space to itself.  Moving forward, we will focus on $G_{AA}(\tau,\tau')$, drop the subscript, and simply refer
to $G(\tau,\tau')$.

As a consequence, for time points $(\tau_1, \tau_2 \cdots , \tau_N)$ the Gramian matrix $\underbar H_{ij}=G(\tau_i,\tau_j)$ is positive semidefinite.
Moreover, if the time points are uniformly spaced at distance $\Delta\tau$, all anti-diagonal entries of $\underbar H$ are identical and $\underbar H$ has the structure of a Hankel matrix.
\begin{figure*}[t]
    \centering\vspace{-7cm}
    \includegraphics[width=\textwidth]{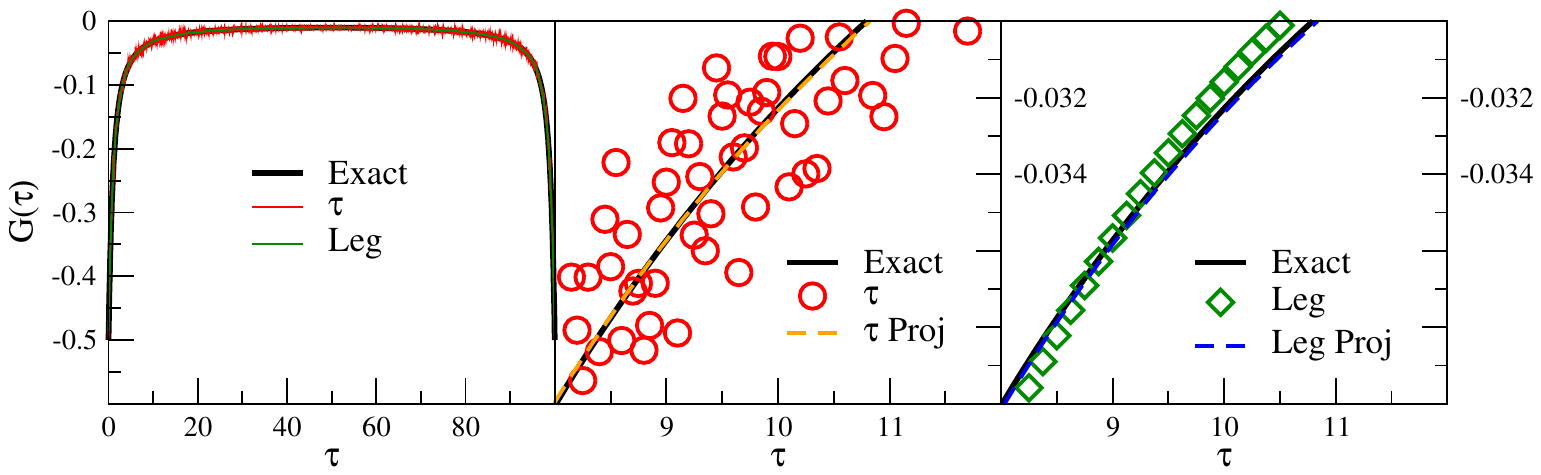}\vspace{-1.5cm}
    \caption{Left panel: Imaginary time Green's function corresponding to a semicircular density of states at $U=0$ and $\beta=100$ showing the exact data, data obtained with  hybridization expansion continuous-time quantum Monte Carlo (CT-HYB) using the imaginary time estimator of Ref.~\cite{Werner06} (`$\tau$') and the Legendre estimator of Ref.~\cite{Boehnke11} (`Leg'). Middle panel: Zoom with $\tau$ data and projection. Right panel: Zoom with Legendre data and projection.
    }
    \label{fig:CTHYBU0}
\end{figure*}
In that case, the Gramian matrix has entries
$H_{ij}=G_{i+j}=G(\tau_i+\tau_j)$ with $\tau_i=i \Delta\tau$, $i=0, \cdots, N/2=\beta/(2 \Delta\tau)$  ($N$ even). This matrix is illustrated in Fig.~\ref{fig:Hankel}, and by the argument above it must be PSD. In addition, its submatrix with the first column and last row dropped (depicted with dashed lines in Fig.~\ref{fig:Hankel}), which corresponds to time points $(\frac{\Delta\tau}{2},\frac{3\Delta\tau}{2}, \cdots,\frac{(N-1)\Delta\tau}{2} )$, is also PSD.

\section{Algorithm}
A Hankel matrix $\tilde{\underbar{H}}_{ij}$ obtained from approximate or statistical simulations of a correlation function $\tilde{G}$ is typically not PSD due to approximation errors or noise. It is therefore natural to construct the PSD Hankel matrix $\underbar{H}_{ij}$ that is closest to $\tilde{\underbar{H}}_{ij}$ within an appropriately
chosen norm, from which we can extract a denoised $G$. Here, we choose the Frobenius norm.
Additionally, $G$ should satisfy physical constraints, such as fermion or boson (anti)-commutation relations or predetermined values at $0$ and $\beta$. 

\subsection{Projection algorithm}\label{sec:projalg}
Before outlining the procedure for finding $G$, we make the following
observations regarding PSD and Hankel matrices.
\begin{enumerate}
    \item[(i)] In the Frobenius norm,
the closest PSD matrix to a given Hankel matrix is obtained by diagonalizing the matrix and setting all negative eigenvalues to zero~\cite{Higham88}. The resulting matrix is in general not a Hankel matrix. PSD matrices form a convex set, since for any two PSD matrices $P, Q$ and any $0\leq\alpha\leq1$ the matrix $A=\alpha P+(1-\alpha)Q$ is also PSD;
this follows straightforwardly from the observation that $x^T Ax = \alpha x^T Px + (1-\alpha)x^TQx \geq 0$ for any vector $x$.

\item[(ii)] The closest Hankel matrix (in the Frobenius norm) to any positive matrix is obtained by averaging the antidiagonals)~\cite{Eberle03}. The resulting matrix is in general not PSD. Hankel matrices also form a convex set since, for two Hankel matrices $H_1$ and $H_2$, $H_3=\alpha H_1+(1-\alpha)H_2$ is also a Hankel matrix.

\item[(iii)] Values of correlation functions at $\tau=0$ or $\tau=\beta$ are often known to much higher precision than values at arbitrary times since they correspond to static expectation values. The $M\times M$ matrix $\underbar B$ ($M=\frac{N}{2}+1)$ with fixed values at the $(0,0)$ and/or $(M, M)$ position closest to a given matrix $\underbar A$ is obtained simply by replacing the values of $\underbar A$ at $(0,0)$ and $(M, M)$ by the desired values. If $\underbar A$ is a PSD Hankel matrix, $\underbar B$ retains the Hankel structure but is in general not PSD. The set of matrices with fixed values at $(0,0)$ and $(M, M)$ is convex. 
\end{enumerate}
At least one positive definite Hankel matrix with a given value at $\tau=0$ and $\beta$ exists,
since the exact solution to the problem satisfies these properties. Since the intersection of these convex sets is not empty, it is possible to use Dykstra's algorithm~\cite{Dykstra86} (see also Appendix~\ref{sec:Dykstra}) to find the intersection point of these convex sets closest to any given set of noisy data. Dykstra's algorithm performs repeatedly projections onto PSD matrices, onto the two Hankel matrices of Fig.~\ref{fig:Hankel}, and onto matrices with fixed values at $(0,0)$ and $\left(M, M\right)$ until convergence is achieved within a predefined tolerance.
The PSD projections may enforce both the PSD structure of $\underbar H$ and the PSD structure of the submatrix illustrated in Fig.~\ref{fig:Hankel}.
While the convergence of the algorithm may be slow, for typical numbers of time slices (50-5000) the numerical effort remains negligible compared to the QMC simulation cost for obtaining data. Projections typically take less than a second per iteration on a single core.
We provide an implementation of the algorithm in Ref.~\cite{repolink}.

\section{Results}
\subsection{QMC data}
We now discuss results of applying the denoising procedure of Sec.~\ref{sec:projalg} to the example of quantum Monte Carlo data. We first consider data obtained with the numerically exact hybridization expansion QMC method~\cite{Werner06,Gull11RMP}.
Methods of this type form the backbone of modern real-materials calculations within DFT+DMFT \cite{Kotliar06} and are used for understanding minimal models of strongly correlated quantum many-body systems \cite{Qin22}. Numerous generalizations~\cite{Werner06Kondo,Gull11RMP}, improvements, and open source implementations exist~\cite{Bauer11,Seth16,Shinaoka17,Wallerberger19}.

We apply the method to the fermionic imaginary time
Green's function of the non-interacting $(U=0)$ half-filled quantum impurity system coupled to a `semicircular' density of states (for details see App.~\ref{app:impurity_model}).
We calculate 
\begin{align}
    G(\tau) = -\text{Tr}\left[\rho c(\tau) c^\dagger(0) \right].
\end{align}
This system corresponds to the exact solution of a Bethe lattice model in the infinite coordination number limit  \cite{Metzner1989, MullerHartmann89} and is frequently studied in the context of dynamical mean field theory \cite{georges1996dynamical}. While the solution for $U=0$ is available analytically, its simulation within the hybridization expansion requires the statistical sampling of a large number of terms in a diagrammatic expansion.

In the hybridization expansion, $G(\tau)$ can either be estimated using a binning method in imaginary time (we designate this as the ``$\tau$'' estimator~\cite{Werner06}) or by expanding the solution into orthogonal polynomials and sampling their coefficients (we designate this the ``Legendre'' estimator~\cite{Boehnke11}). 

Fig.~\ref{fig:CTHYBU0} shows the improvement that can be obtained
by making use of the projection for this model.
Results and error bars are obtained from 64 independent Monte Carlo runs.
The left panel shows an imaginary-time Green's function obtained at temperature $T=1/100$ (we set the hopping parameter to $t=1$). Data measured with both estimators is consistent with the exact solution within Monte Carlo error bars (not shown) but Monte Carlo noise is clearly visible.
The middle panel shows the same $\tau$ estimator data, focused on $8\leq \tau \leq 12$. The result from the projection of this noisy data is shown as an orange dashed line and is consistent with the exact solution.

The right panel shows the same region for the Legendre estimator. In this case, because the Legendre expansion introduces correlations in imaginary time, the data appears smooth but nevertheless shows considerable deviation from the exact result.
Similarly, this deviation is eliminated in the projected data that (within the resolution of the plot) coincides with the exact data.
\begin{figure}
    \centering
    \includegraphics[width=\columnwidth]{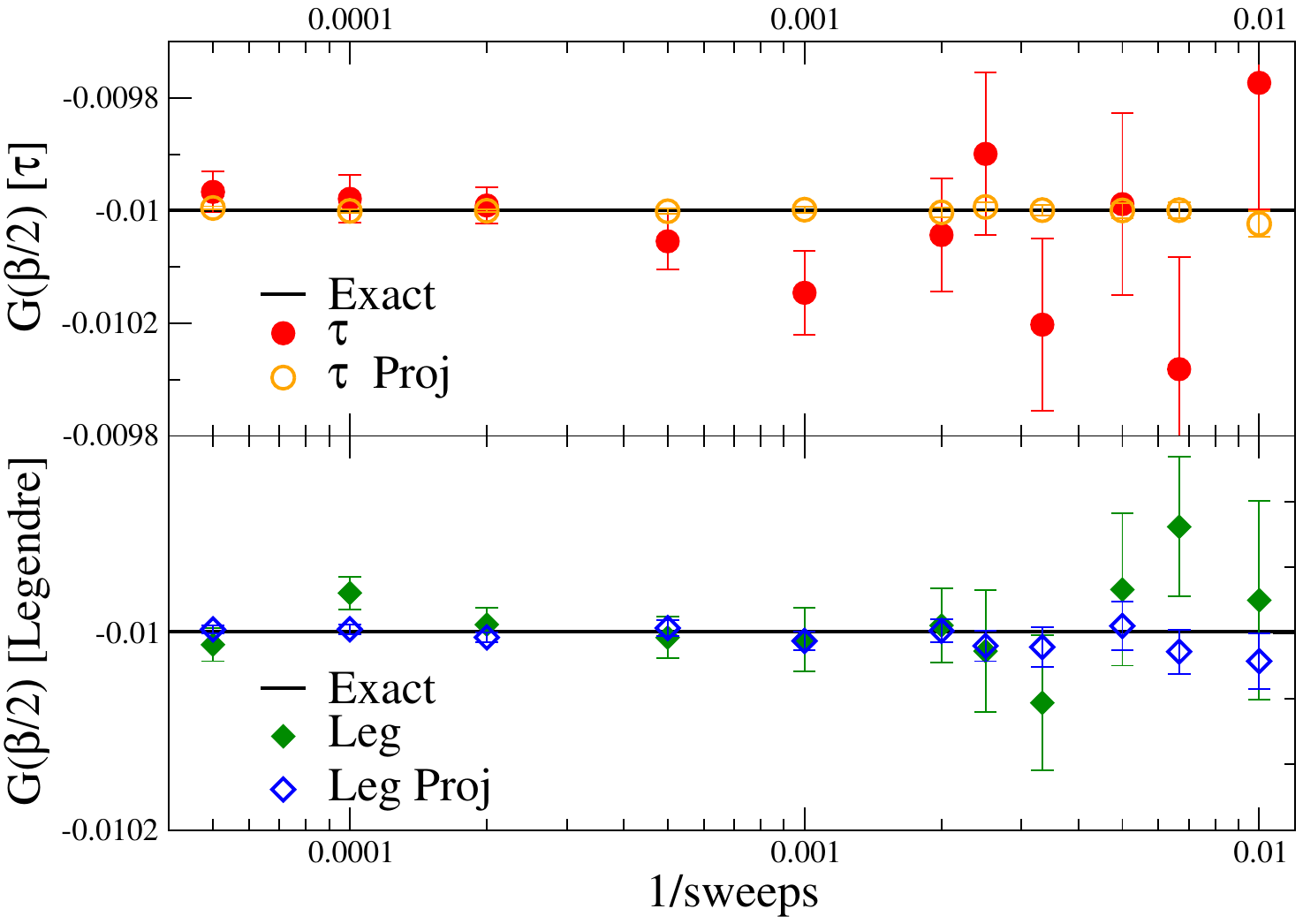}\vspace{-0.75cm}
    \caption{Monte Carlo error of system of Fig.~\ref{fig:CTHYBU0} as a function of sweeps obtained from $64$ independent runs. Black and red, top panel: $\tau$ estimator of Ref.~\cite{Werner06}
 and projection plotted with error bars. Green and blue, bottom panel: Legendre measurement of Ref.~\cite{Boehnke11}. Note the different $y$-axis scales.
    \label{fig:ErrorScaling}}
\end{figure}

In Fig.~\ref{fig:ErrorScaling}, we present a detailed error analysis for the same system, focusing on $G(\beta/2)$ which, at low $T$, is related to the density of states at the Fermi energy \cite{Trivedi95}. We show the result from $64$ independent evaluations of the two estimators;
as expected from a numerically exact method, when the number of sweeps increases, the estimators
approach the exact result. 

We next project each of the $64$ samples and average the projected Green's functions. For both estimators, the projected data are consistent with the exact result within error bars for all samples, confirming that the projection did not introduce a bias. 
For all numbers of sweeps considered, the error bars on the projected $G(\beta/2)$ using the $\tau$ estimator are approximately $20$ times smaller than those before projection, corresponding to a $20^2=400$-fold saving in CPU time. 
Similarly, for the Legendre estimator, the average error is reduced by around a factor of five (corresponding to a 25-fold saving in CPU time).
 For both estimators, this factor remains approximately constant as the number of sweeps is increased.

We note that the projection employed here is highly non-linear and may in general deform a Gaussian distribution, with details depending on the type of Monte Carlo algorithm and the observable estimator. It is therefore dangerous to perform the projection with just one noisy sample (such as the mean value of a QMC calculation). Rather, a careful statistical analysis using, for example, jackknife or bootstrap error propagation of independent data is advised. In this paper, we first perform the Hankel projection on multiple independent data sets and only then average the results.

 \begin{figure}
 \centering
    \includegraphics[width=\columnwidth]{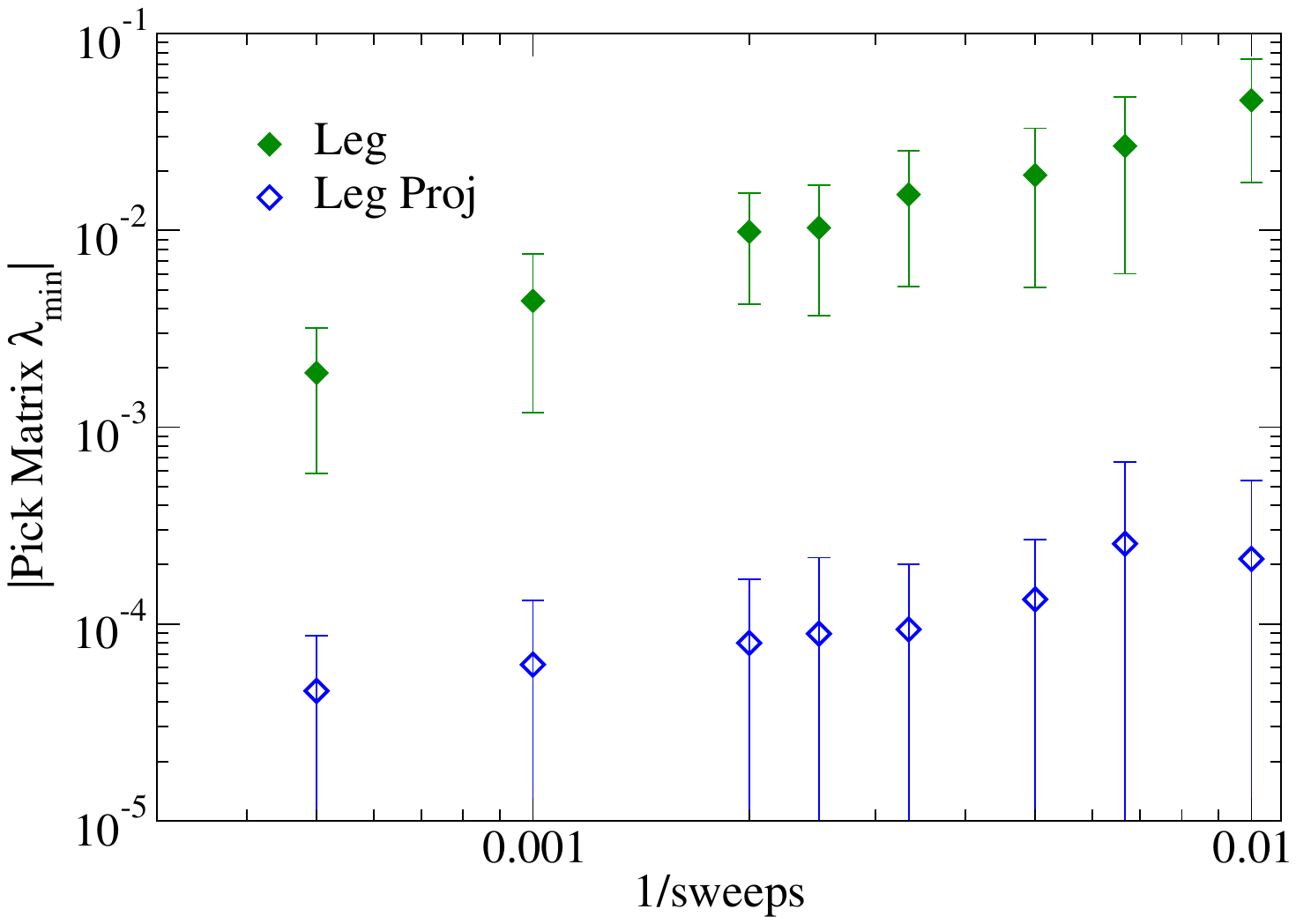}\vspace{-0.75cm}
\caption{Average absolute value and error of the lowest eigenvalue of the Pick matrix, calculated from 64 independent Monte Carlo runs, as a function of the inverse number of sweeps. Legendre measurement. Green: Original data. Blue: Data after projection. System as in Fig.~\ref{fig:CTHYBU0}.}
    \label{fig:Pick}
\end{figure}

\subsection{Positive spectral functions}
\begin{figure*}
    \centering
    \vspace{-8cm}
    \includegraphics[width=\textwidth]{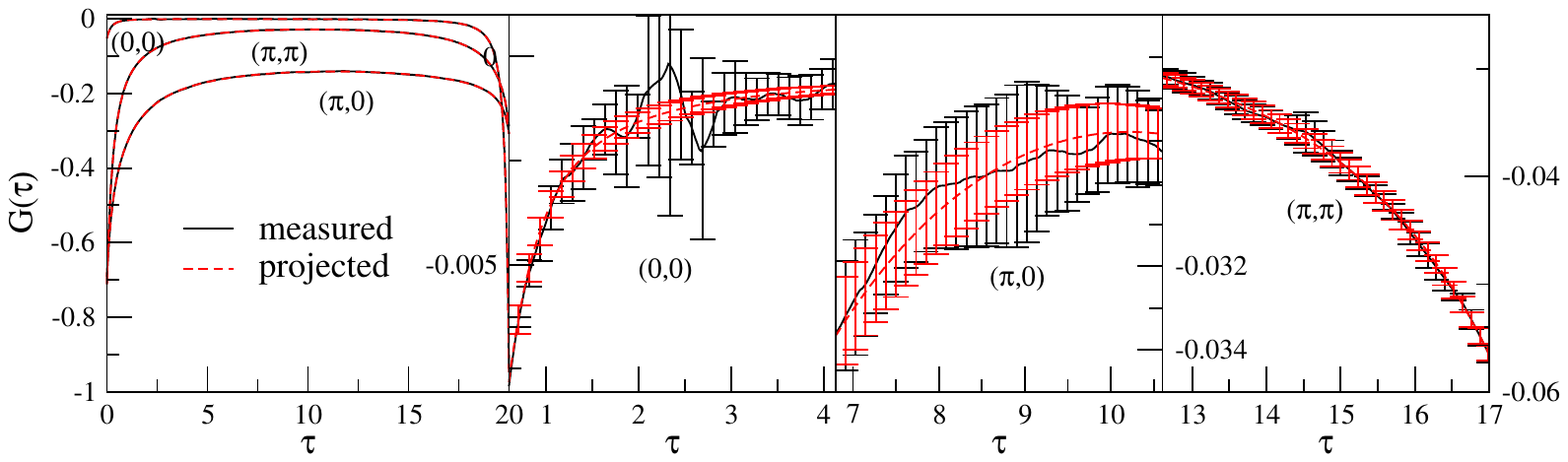}\vspace{-0.75cm}\vspace{-0.5cm}
    \caption{Projection of a strongly frustrated four-site DCA data cluster \cite{maier2005quantum} as obtained from an auxiliary field continuous-time quantum Monte Carlo (CT-AUX)~\cite{Gull2008} simulation (black) and Hankel projection (red). Left panel: Green's functions at $(0,0)$, $(\pi,0)$, and $(\pi,\pi)$. Middle panel: zoom for $(0,0)$ Green's function. Right panel: zoom for $(\pi,0)$ Green's function.
    } 
    \label{fig:ClusterError}
\end{figure*}

Is there a positive spectral function that corresponds to a set of imaginary time points? This question is answered by the Pick criterion \cite{Pick17}, which states that if the so-called Pick matrix
for Matsubara data is formed, a positive spectral function consistent with the data only exists if the matrix has no negative eigenvalues \cite{fei2021nevanlinna}. See App.~\ref{app:pick} for a brief discussion.

The exact solution of a causal quantum problem has a Pick matrix with only non-negative eigenvalues. However, Monte Carlo noise generally introduces negative eigenvalues to the Pick matrix.
Fig.~\ref{fig:Pick} shows the average of the absolute value of the minimal ({\it i.e.} the most negative) eigenvalue $\lambda_\text{min}$ of the Pick matrix constructed from the Legendre data of Fig.~\ref{fig:CTHYBU0} (green) and its projection (blue). 
The projection reduces the magnitude of $\left|\lambda_\text{min}\right|$ by 1-2 orders of magnitude on average, demonstrating that the projected data is much closer to a positive spectral function than the original results. 
The remaining negative contribution may stem from a combination of systematic errors such as the projection convergence cutoff and Fourier transform precision artifacts, as well as from errors that are consistent with the inner product introduced here but inconsistent with causality.
Aside from ensuring that the Green's functions are physical, 
the existence of a positive spectral function is a prerequisite for obtaining precise analytic continuations with modern methods~\cite{fei2021nevanlinna,Fei2021Caratheodory,Ying22,Zhen23,Zhang23}.

\subsection{Interacting model}
The projection method is independent of the algorithm and the system studied and is particularly useful in systems with strong noise, including systems with a sign problem caused by, for example, frustration. The square-lattice Hubbard model with a large next-nearest hopping term $t'=t=1$ is such a frustrated model. Fig.~\ref{fig:ClusterError} shows the Green's function of a frustrated four-site cluster at temperature $T=1/20$, chemical potential $\mu=0$, and interaction $U=5$, coupled to a non-interacting bath (see App.~\ref{app:hubbard_model}). Systems of this type appear in the simulation of ``dynamical cluster approximation'' (DCA) cluster dynamical mean field problems \cite{maier2005quantum}. The system is solved with a continuous-time~\cite{Gull11RMP} auxiliary field \cite{Gull2008} quantum impurity solver, using a Green's function estimator formulated in Matsubara frequency space \cite{Rubtsov2005}. The average sign of the problem is $\sim0.63$.

The left panel of Fig.~\ref{fig:ClusterError} shows the (strictly negative) momentum-space Greens' functions at the three independent momentum points $k=(0,0)$, $k=(0,\pi)$ and $k=(\pi,\pi).$  The middle panel shows measured data along with the projection for $k=(0,0)$, and the right panel for $k=(\pi,0).$
Error bars are shown on every $5$th data point.

The improvement of the projection over the measured data is around a factor of two on average (corresponding to a four-fold saving of computer time). The smaller improvement may be due to the different formulation of the Monte Carlo estimator (which measures the Fourier transform of the data shown \cite{Rubtsov2005} and therefore already induces strong correlations between time points). Note that the smaller advantage observed here is not due to the sign problem. We have observed cases (not shown) where the Hankel projection offers greater benefits as the temperature is lower and the sign problem becomes more severe.

\section{Discussion}
The Hankel projection introduced here is a generic post-processing method that can be applied to any imaginary-time response function data at negligible additional cost. As we have shown, it removes unphysical components of the noise and thereby leads to data that is much more precise. A careful error analysis has shown that no bias is introduced by the projection, and that projected data is substantially closer to a ``causal'' solution than unprojected data. Hankel projections of this type should therefore be applicable in all imaginary-time and Matsubara frequency calculations where noise (such as stochastic noise, measurement noise, or approximation errors) is a source of error.

The advantage of the method strongly depends on the type of estimator and Monte Carlo algorithm used. In continuous-time algorithms, where very many independent time points are measured, the method is more effective than in discrete-time setups~\cite{Blankenbecler1982,Hirsch1986} where typically only a few correlated imaginary-time slices are measured. 
Interaction-~\cite{Rubtsov2005,Gull2008} and hybridization expansion \cite{Werner06} methods employ different estimators with entirely different error structures and therefore lead to different improvements of the result. Applications to diagrammatic Monte Carlo of various types, such as inchworm Monte Carlo, bold-line Monte Carlo, to matrix-valued correlators, and to bosonic systems \cite{Pollet12} will be interesting to explore.

Similarly, it will be interesting to explore if this method can remove unphysical components of Green's functions obtained with methods with systematic approximations, such as certain non-causal vertex-corrected beyond-GW methods~\cite{Stefanucci14} or tensor-train approximations \cite{Yuriel22,Andre23}. In those methods, there is no stochastic `noise' component but a systematic approximation error that may violate causality, which is closely related to the inner product structure discussed in this paper~\cite{Kemper23}.

\begin{acknowledgments}
We thank Andre Erpenbeck and Edwin Huang for useful comments.
YY was sponsored by NSF QIS 2310182. AFK was supported by the Department of Energy, Office of Basic Energy Sciences, Division of Materials Sciences and Engineering under grant no. DE-SC0023231.  
EG and CY were supported by the  U.S. Department of Energy, Office of Science, Office of Advanced Scientific Computing Research and Office of Basic Energy Science, Scientific Discovery through Advanced Computing (SciDAC) program under Award Number(s) DE-SC0022088.
The data for the figures are submitted as a supplement to this paper.
\end{acknowledgments}

\bibliography{refs}

\appendix

\renewcommand\thefigure{S\arabic{figure}}  
\renewcommand\thetable{S\arabic{table}}  
\setcounter{figure}{0}

\section{Dykstra's projection algorithm}\label{sec:Dykstra}

Dykstra’s algorithm~\cite{Dykstra86} is a method that computes a point in the intersection of convex sets. For a specified initial point, Dykstra’s algorithm will find the projection of that initial point onto the intersection. In other words, Dykstra’s algorithm will find the point in the intersection that is closest to the initial point.

Suppose we have two convex sets $A$ and $B$ which we know we can do projections onto them via operation $\mathcal{P}_A$ and $\mathcal{P}_B$, and we want to project the initial point $x_0$ on the intersection of $A$ and $B$. We set another two auxiliary variables $a$, $b$ to store the intermediate updates, where the initial values are $a_0=b_0=0$. Dykstra’s algorithm updates the values of these parameters at step $k+1$ according to their value at step $k$ via
\begin{subequations}
\begin{align}
& x_{k+1}^{(a)}=\mathcal{P}_A\left(x_k-a_k\right) \label{eqn:Dykstra_1},\\
& a_{k+1}=x_{k+1}^{(a)}-(x_k-a_k) \label{eqn:Dykstra_2},\\
& x_{k+1}^{(b)}=\mathcal{P}_B\left(x_{k+1}^{(a)}-b_k\right) \label{eqn:Dykstra_3},\\
& b_{k+1}=x_{k+1}^{(b)}-(x_{k+1}^{(a)}-b_k) \label{eqn:Dykstra_4},\\
& x_{k+1}=x_{k+1}^{(b)}.
\end{align}
\end{subequations}
By the end of the calculation, the variables x converge to the intended point of projection. It is obvious that the updates from $\mathcal{P}_A$ (Eqs.~\ref{eqn:Dykstra_1} and \ref{eqn:Dykstra_2}) are identical to those from $\mathcal{P}_B$ (Eqs.~\ref{eqn:Dykstra_3} and \ref{eqn:Dykstra_4}), except for their inputs. This allows us to easily adapt the same approach when applying Dykstra’s algorithm to more than two convex sets.

In applying Dykstra’s algorithm to Hankel projections, we start with an initial Hankel matrix created from a given Green’s function. We have four convex sets as detailed in Sections \ref{sec:method} and \ref{sec:projalg}. The corresponding projections are:
\begin{itemize}
    \item Projection onto positive semidefinite (PSD) matrices.
    \item Projection onto matrices with specific values at $G(0)$ and $G(\beta)$.
    \item Projection onto the entire Hankel matrix, as shown in Fig.~\ref{fig:Hankel}.
    \item Projection onto the small Hankel matrix, outlined with a dashed line in Fig.~\ref{fig:Hankel}.
\end{itemize}

These four projections are performed sequentially in a single update of Dykstra’s algorithm, similar to projection $\mathcal{P}_A$ and $\mathcal{P}_B$ described above. The process continues until the Hankel matrices from two consecutive iterations are close enough, or until the calculation reaches a maximum number of iterations. At the end of the calculation, the projected Green’s function can be directly read from the converged Hankel matrix.

\section{Single impurity coupled to an infinite-dimensional Bethe lattice }
\label{app:impurity_model}
We study an impurity of two spin degrees of freedom coupled to a bath with a dispersion characterized by a semi-elliptical form having a bandwidth of $4t$. The corresponding density of states of the bath is given by $D_{\uparrow\uparrow}(\omega)=D_{\downarrow\downarrow}(\omega)=\frac{1}{2\pi t^2} \sqrt{4 t^2-\omega^2}$ for $-2 t \leq \omega \leq 2 t$, and the hybridization function is expressed as $\Delta_{\uparrow\uparrow}(\tau)=\Delta_{\downarrow\downarrow}(\tau)=-\int d \omega D(\omega) \frac{e^{-\tau \omega}}{1+e^{-\beta \omega}}$ for $0\leq \tau \leq \beta$. The action of this impurity reads
\begin{equation}
\mathcal{S}_{\text{imp}} = \mathcal{S}_{\text{loc}} + \int_{0}^{\beta} d \tau \int_{0}^{\beta} d \tau^{\prime} \sum\limits_{\sigma={\uparrow, \downarrow}} c^{*}_{\sigma}(\tau) \Delta_{\sigma \sigma}(\tau-\tau^{\prime}) c_{\sigma}(\tau^{\prime}),  
\end{equation}
\begin{equation}
    \begin{aligned}
\mathcal{S}_{\text{loc}}
&= \int_{0}^{\beta} d \tau \sum\limits_{\sigma={\uparrow, \downarrow}} c^{*}_{\sigma}(\tau) (\partial_{\tau} - \mu)c_{\sigma}(\tau) \\ 
&+ U\int_{0}^{\beta}d \tau c^{*}_{\uparrow}(\tau){c}_{\uparrow}(\tau)c^{*}_{\downarrow}(\tau){c}_{\downarrow}(\tau).
\end{aligned}
\end{equation}
The hybridization expansion \cite{Werner06} treats $\mathcal{S}_{\text{loc}}$ exactly and performs a perturbation expansion of the hybridization term around this solution.

The dynamical mean field formalism is reviewed in Ref.~\cite{georges1996dynamical}.

\section{DCA Model Hamiltonian}
\label{app:hubbard_model}
The DCA data of Fig.~\ref{fig:ClusterError} are obtained on a four-site cluster approximation to the Hubbard model \cite{Jarrell98,Jarrell01,maier2005quantum} on a heavily frustrated square lattice with $t'=t=1$ as in \cite{Li20}. Shown are the results for the first iteration at an interaction strength of $U=5$, starting from the DCA solution of the non-interacting model.

The choice of parameters is motivated by the fact that the frustration $t=t'$ leads to a strong fermion sign problem that drastically amplifies noise. While the projection cannot overcome the exponential cost of the sign problem, it alleviates the size of the noise and therefore the prefactor of the exponential scaling with temperature or interaction.

The DCA formalism is reviewed in Ref.~\cite{maier2005quantum}.

\section{Pick criterion}
\label{app:pick}
In this appendix, we briefly describe the Pick criterion for
Matsubara Green's functions. The criterion is based on work by
Pick \cite{Pick17}, and was introduced to the context of fermionic Matsubara
Green's functions in Ref.~\cite{fei2021nevanlinna}.

The problem of analytic continuation is to find a function $\mathcal{G}$
\begin{align}
    \mathcal{G}: \mathbb{C} \rightarrow \mathbb{C}
\end{align}
that is analytic on
the upper half of the complex plane,  $\mathbb{C}^+$, and coincides with the
available numerical data $G$ on the Matsubara axis, i.e. $G_n = \mathcal{G}(z=i\omega_n)$.
Because $\mathcal{G}$ is
to represent a Green's function, it should have an imaginary part
with  a definite sign. Such a function  is called a
Nevanlinna function.

A priori it is not known whether a Nevanlinna function exists that
passes through numerical data with stochastic or systematic errors.
This question, and thus the
question of whether a physically correct  Green's function (i.e. analytic on
$\mathbb{C}^+$ and with appropriate sign of the spectrum) exists,
is answered by the Pick criterion \cite{Pick17}. First,
one maps the (closed) upper half complex plane $\overline{\mathbb{C}^+}$ 
to the (closed) complex unit disk $\overline{\mathbb{D}}$ via
a M\"obius transformation:
\begin{align}
    &h:\ \overline{\mathbb{C}^+} \rightarrow \overline{\mathbb{D}}\\
    &h(z) := \frac{z-i}{z+i}.
\end{align}
Next, one forms the Pick matrix out of the numerical data
$G_n$ and the M\"obius-transformed Matsubara frequencies
$h(\omega_n)$:
\begin{align}
    \mathcal{P}_{n,m} = \left[
        \frac{G_n + G_m^*}{1-h(\omega_n) h(\omega_m)^*}
    \right],\quad
    n,m = 1,2,\ldots,M.
\end{align}
Note that the transformation is not strictly
necessary, and that a similar matrix can be formulated
without the M\"obius transformation~\cite{Fei2021Caratheodory}.
The Pick criterion states that a Nevanlinna interpolant
exists if and only if $\mathcal{P}$ is positive semidefinite (PSD),
and that a unique solution exists if and only if $\mathcal{P}$ is singular.
In other words, evaluating the degree to which $\mathcal{P}$ for a given
numerical Green's function $G$ fails to
be PSD is a measure of the causality violation of a numerical Green's function.

\end{document}